\documentclass[aps,pre,floats,floatfix,twocolumn,showpacs,superscriptaddress,nofootinbib]{revtex4-1}
\usepackage{latexsym,amsmath}
\usepackage{amsbsy}
\usepackage[usenames]{color}
\usepackage[pdftex]{graphicx}
\usepackage{amssymb}
\usepackage{epstopdf}

\begin{document}

\title{Hidden Variables in Bipartite Networks}
\date{\today}
\author{Maksim Kitsak}
\affiliation{Cooperative Association for Internet Data Analysis (CAIDA), University of California, San Diego (UCSD),
9500 Gilman Drive, La Jolla, CA 92093, USA}

\author{Dmitri Krioukov}
\affiliation{Cooperative Association for Internet Data Analysis (CAIDA), University of California, San Diego (UCSD),
9500 Gilman Drive, La Jolla, CA 92093, USA}

\begin{abstract}
We introduce and study random bipartite networks with hidden variables. Nodes in these networks are
characterized by hidden variables which control the appearance of links between node pairs. We derive analytic
expressions for the degree distribution, degree correlations, the distribution of the number of common neighbors, and the
bipartite clustering coefficient in these networks. We also establish the relationship between degrees of nodes in original
bipartite networks and in their unipartite projections. We further demonstrate how hidden variable formalism can
be applied to analyze topological properties of networks in certain bipartite network models, and verify our analytical results in
numerical simulations.
\end{abstract}

\maketitle
\section{Introduction}
\begin{figure}
\includegraphics[width=8.0 cm,angle=0]{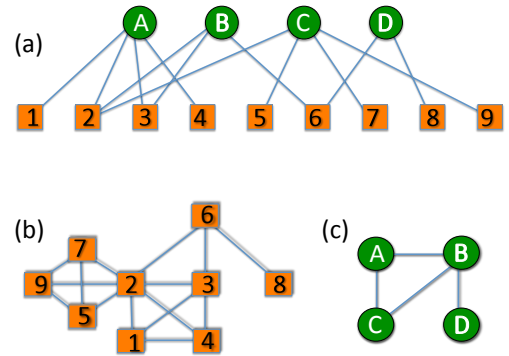}
\caption{ \footnotesize (Color Online) \label{fig:bip} A toy bipartite network and its unipartite projections.
(a)~Original bipartite network. We refer to the nodes of one type as top nodes (labeled by letters) and to the nodes of
the other type as bottom nodes (labeled by numbers). Unipartite projections of the original network onto (b)~bottom and
(c)~top domains. The top (bottom) nodes are connected in the projections if they have at least one common
neighbor in the original network.}
\end{figure}

Bipartite networks are composed of two types of nodes with no links connecting nodes of the same type, see
Fig.~\ref{fig:bip}(a).
Examples include recommendation systems~\cite{uchyigit08}, networks of collaborations~\cite{ramasco04} and metabolic
reactions~\cite{ma03}, gene regulatory networks~\cite{davidson05}, peer to peer networks~\cite{iamnitchi03},
pollination networks~\cite{burgos08}, and many others~\cite{latapi08}. Compared to traditional unipartite networks,
less is known about the organizing principles determining the structure and evolution of bipartite networks, partly
because only unipartite projections of bipartite networks are often considered. The unipartite projection accounts for
connecting two nodes of one type by a link if these nodes share at least one neighbor of the other type, and then
throwing out all nodes of this other type, see Figs.~\ref{fig:bip}(b) and \ref{fig:bip}(c).
Even though this procedure allows one to study bipartite networks using powerful tools developed for unipartite
networks, the unipartite projections in most cases lead to significant loss of information, and to artificial inflation
of the projected network with fully connected subgraphs~\cite{latapi08,guillaume04}.

Nodes in real bipartite networks can often be characterized by a number of intrinsic attributes.
For example, in
recommendation networks, composed of consumer and product nodes, a consumer-product pair is connected if
the consumer has purchased the product. Consumers can be characterized by their age,
geographic location, income, sex, lifestyle, etc., while products have their type, price, quality,
uniqueness, and other properties. Consumers do not buy products at random.
Making their purchase decisions, consumers implicitly match their attributes with those of products.
For example, a person
with a higher income is more likely to purchase an expensive item, books in Italian are mostly purchased by people who
speak Italian, consumers at a gas station tend to own a car, etc. Similar considerations apply to the
formation of links between researchers and scientific projects, molecules and reactions in which they participate,
and so forth.

The concept of hidden variables formalizes these observations as follows. Every node of each type in a bipartite
network is assigned a number of hidden variables drawn from some distributions, and then every node pair of different
types is connected with some probability which depends on the hidden variables of the two nodes. In this work we build
the hidden variable formalism for bipartite networks, based on the formalism developed earlier for unipartite
networks~\cite{boguna03}. Specifically, in Section~II we overview basic topological characteristics of bipartite
networks. In Section~III we define a general class of bipartite networks with hidden variables, and study analytically
the topological properties of networks in this class. In Section~IV we consider two specific examples of bipartite
networks with hidden variables, uncorrelated and stratified bipartite networks, and confirm in simulations our
analytical results for these networks. Section~V summarizes the paper.

\section{Topological Characteristics of Bipartite Networks}

In this section we review some key relationships among the basic topological characteristics of bipartite networks.

Let the nodes of two different types be called top and bottom nodes, see Figs.~\ref{fig:bip}(b) and
\ref{fig:bip}(c).
Similar to unipartite networks, the degree correlations in bipartite networks are defined by the number of links
$E_{k\ell}$ between top and bottom nodes of degrees $k$ and
$\ell$~\cite{callaway01}. The correlation matrix $E_{k\ell}$ satisfies the following equations:
\begin{equation}
\label{eq:identity} \sum_{\ell} E_{k\ell} = k N_{k},\quad \sum_{k} E_{k\ell} = \ell M_{\ell},\quad \sum_{k, \ell}
E_{k\ell} = E,
\end{equation}
where $N_{k}$ and $M_{\ell}$ are the numbers of top and bottom nodes of degree $k$ and $\ell$, and $E$ is the total
number of links in the network. The joint
degree distribution $P(k,\ell)$ is the normalized correlation matrix, i.e.,
the probability that a randomly chosen edge connects nodes of degrees $k$ and $\ell$:
\begin{equation}
\label{eq:cond_joint} P(k,\ell) = {E_{k\ell} \over E},
\end{equation}
which contains all information needed to construct a network with a given degree
distribution and correlations.

The top and bottom node degree distributions $P(k)$ and $P(\ell)$ can be obtained from
Eq.~(\ref{eq:identity}):
\begin{equation}
\label{eq:distributions}
P(k) = {\overline{k} \over k} \sum_{\ell} P(k,\ell),\quad
P(\ell) = {\overline{\ell} \over \ell} \sum_{k} P(k,\ell).
\end{equation}
The conditional probabilities $P(\ell|k)$ and $P(k|\ell)$ that an edge emanating from a $k$- or $\ell$-degree node is connected to a node of degree $\ell$ or $k$ are
\begin{eqnarray}
\label{eq:cond_prob1} P(\ell|k) &=& {E_{k\ell} \over k N_{k}} = {\overline{k} P(k,\ell) \over k P(k)},\\
\label{eq:cond_prob2} P(k|\ell) &=& {E_{k\ell} \over \ell N_{\ell}} = {\overline{\ell} P(k,\ell) \over \ell P(\ell)}.
\end{eqnarray}

To characterize degree correlations in unipartite networks, one often considers the average nearest neighbor
degree (ANND), which is the average degree of the neighbors of all $k$-degree nodes~\cite{satorras01}.
The ANNDs for top and bottom nodes in a bipartite network are
\begin{equation} \label{eq:ANND_def1}
\overline{\ell}_{nn}(k) = \sum_{\ell} \ell P(\ell|k),\quad
\overline{k}_{nn}(\ell) = \sum_{k} k P(k|\ell).
\end{equation}
In uncorrelated bipartite networks
\begin{equation}
P^{unc}(k,\ell) = {k P(k)\over \overline{k}} {\ell P(\ell) \over \overline{\ell}}.
\end{equation}
As a result, $P(\ell|k)$ and $P(k|\ell)$ do not depend on $k$ and $\ell$, respectively:
\begin{equation}
\label{eq:cond_prob1_unc}
P^{unc}(\ell|k) = {\ell \over \overline{\ell}} P(\ell),\quad
P^{unc}(k|\ell) = {k \over \overline{k}} P(k),
\end{equation}
and neither do the ANNDs:
\begin{equation} \label{eq:ANND_def2}
\overline{\ell}^{unc}_{nn}(k) = {\overline{\ell^{2}} \over \overline{\ell}},\quad
\overline{k}^{unc}_{nn}(\ell) = {\overline{k^{2}} \over \overline{k}}.
\end{equation}
Networks with increasing or decreasing ANNDs are called assortative or disassortative~\cite{newman01}.
Some real bipartite networks have non-trivial degree correlation profiles, and therefore they can not be
classified as either assortative or disassortative~\cite{latapi08}.

The standard clustering coefficient of node $i$ quantifies how close $i$'s neighbors are to forming a clique~\cite{watts98}:
\begin{equation}
c(i) = {2 \over k_i (k_i - 1)} \sum_{j > k} e_{jk},
\end{equation}
where the summation is over all $i$'s pairs of neighbors $j$ and $k$, and $e_{jk}$ is the adjacency matrix. Since in
bipartite networks there are no loops of size $3$, this clustering coefficient is zero for all nodes. Therefore, to
assess the density of connections in a vicinity of a particular node, one has to analyze connections among its second
nearest neighbors. There have been several attempts to generalize the clustering coefficient for bipartite networks
using this idea~\cite{latapi08,lind05,zhang08}. Here we focus on the definition by Zhang et al~\cite{zhang08}:
\begin{equation}
c_B(i) = {\sum_{m > n} q_{imn} \over \sum_{m > n} \left(q_{imn} + k_m + k_n - 2 \eta_{imn} \right)},
\label{eq:clustering}
\end{equation}
where $\sum_{m > n}$ goes over all pairs of $i$'s neighbors, $q_{imn}$ is the number of common neighbors between nodes
$m$ and $n$ excluding $i$, $k_{m}$ and $k_{n}$ are the degrees of nodes $m$ and $n$, and $\eta_{imn} = 1 + q_{imn} +
e_{mn}$. The above definition may look cumbersome, but it has a simple interpretation. Let $A_{m}$ and $A_{n}$ be the
sets of neighbors of nodes $m$ and $n$ excluding $i$. Then $q_{imn}$ is the intersection of $A_{m}$ and $A_{n}$,
$q_{imn} = \|A_{m} \bigcap A_{n}\|$, while $q_{imn} + (k_m - \eta_{imn}) + (k_n - \eta_{imn})= \|A_{m} \bigcup A_{n}\|$
is their union. Therefore, the bipartite clustering coefficient is simply
\begin{equation}
c_B(i) = {\sum_{m > n} \|A_m \bigcap A_n \| \over \sum_{m > n} \|A_m \bigcup A_n\|}.
\end{equation}
The ratio of the intersection and union of two sets is known as the Jaccard similarity coefficient~\cite{jaccard1901}.
The bipartite clustering coefficient, on the other hand, is given by the ratio of the sums of intersections and unions
for all pairs of $i$'s neighbors. Therefore, the bipartite clustering coefficient can be interpreted as a combined
Jaccard similarity of $i$'s neighbors. Regardless of the clustering definition details, nodes in real bipartite
networks tend to be strongly clustered, as compared to nodes in their randomized counterparts with preserved degree
distributions~\cite{latapi08}.

\section{Hidden Variable Formalism for Bipartite Networks}
We define the class of bipartite networks with hidden variables as follows:
\begin{itemize}
\item[]{(i)} Each top and bottom nodes $i$ and $j$ are assigned hidden variables $\kappa_{i}$ and $\lambda_{j}$ drawn from
probability distribution $\rho_{t}(\kappa)$ and $\rho_{b}(\lambda)$;
\item[]{(ii)} Each top-bottom node pair $\{i,j\}$ is connected with probability $r(\kappa_{i}, \lambda_{j})$, $0 \leq r(\kappa,\lambda) \leq 1$.
\end{itemize}

The hidden variable formalism developed here is valid for both discrete and continuous variables. In the latter case,
all sums must be replaced by integrals. We are primarily interested in the cases where the hidden variable
distributions $\rho_{t}(\kappa)$ and $\rho_{b}(\lambda)$ are independent of the sizes of the top and bottom domains $N$
and $M$. We also assume that in the thermodynamic limit of large $N,M$, these sizes are proportional to each other, $N
\propto M$. For the sake of clarity we consider only one hidden variable per node. The generalization to several hidden
variables per node is straightforward.  We also drop indices in the top and bottom hidden variable distribution
notations: $\rho_{t}(\kappa) \equiv \rho(\kappa)$ and $\rho_{b}(\lambda) \equiv \rho(\lambda)$.

\subsection{Degree distributions}
We first compute the most basic topological properties of the networks in the model---the degree distributions and average degrees.
Due to the stochastic nature of
connections between top and bottom nodes, we can not compute the degree of a top node with hidden
variable $\kappa$ deterministically. Instead, we can compute propagator $g(k|\kappa)$, which is the probability that a
node with hidden variable $\kappa$ ends up connecting to $k$ bottom nodes. Similarly, propagator
$f(\ell|\lambda)$ is the probability that a bottom node with hidden variable $\lambda$ will be connected to $\ell$ top nodes.
Propagators $g(k|\kappa)$ and $f(\ell|\lambda)$ are the main building blocks of the hidden variable formalism. As soon as
we know $g(k|\kappa)$, for example, the average degree $\overline{k}(\kappa)$ of a top node with hidden variable $\kappa$,
the degree distribution $P(k)$, and the average degree $\overline{k}$ in the top node domain are given by:
\begin{eqnarray}
\label{eq:avg_k_kappa} \overline{k}(\kappa) &=& \sum_{k} k g(k|\kappa),\\
\label{eq:pk} P(k) &=& \sum_{\kappa} g(k|\kappa) \rho(\kappa),\\
\label{eq:avg_k} \overline{k} &=& \sum_k k P(k) = \sum_{\kappa} \overline{k}(\kappa) \rho(\kappa),
\end{eqnarray}
while the corresponding expressions for bottom nodes can be obtained by  an appropriate swap of notations.

To compute propagator $g(k|\kappa)$ we first compute partial propagator $g_{i}^{\lambda_i}(k_{i}|\kappa)$ defined
as the probability that a top node with hidden variable $\kappa$ ends up having $k_{i}$ connections to bottom
nodes with hidden variable $\lambda_i$. Since links between node pairs appear independently from one pair to another,
$g_{i}^{\lambda_i}(k_{i}|\kappa)$ is given by the binomial distribution:
\begin{equation}
\label{eq:partial_propagator} g_{i}^{\lambda_i}(k_{i}|\kappa) = C^{M_{\lambda_i}}_{k_{i}}\left[
r(\kappa,\lambda_i)\right]^{k_{i}} \left[1-r(\kappa, \lambda_i) \right]^{M_{\lambda_i}-k_{i}},
\end{equation}
where $C^{a}_{b}$ is the binomial coefficient, and $M_{\lambda_i} \equiv M\rho(\lambda_i)$ is the total number of
bottom nodes with hidden variable $\lambda_i$. The full propagator $g(k|\kappa)$ is then a convolution of partial
propagators:
\begin{equation}
\label{eq:g_propagator} g(k|\kappa) = \sum_{\sum k_{i}=k} \prod_{i} g_{i}^{\lambda_{i}}(k_{i}|\kappa),
\end{equation}
where the product is over the entire spectrum of hidden variables $\lambda$, while the summation
is over the ensemble of all possible degrees $k_{i}$ whose sum is $k$.

Since the full
propagator is a convolution, its generating function $\hat{g}(z|\kappa)$ is a product of the generating functions $\hat{g}^{\lambda}(z|\kappa)$ for
partial propagators:
\begin{eqnarray}
\hat{g}(z|\kappa) &=& \prod_{\lambda}\hat{g}^{\lambda}(z|\kappa),\quad\text{where}\label{eq:g(z|kappa)}\\
\hat{g}(z|\kappa) &\equiv& \sum_{k} g(k|\kappa) z^{k},\\
\hat{g}^{\lambda}(z|\kappa) &\equiv& \sum_{k} g^{\lambda}(k|\kappa) z^{k}.
\end{eqnarray}
The generating function for binomial $g^{\lambda}(k|\kappa)$ is
\begin{equation}
\hat{g}^{\lambda}(z|\kappa) = (1 - z (1-r(\kappa, \lambda)))^{M_{\lambda}},
\end{equation}
substituting which into Eq.~(\ref{eq:g(z|kappa)}) we obtain
\begin{equation}
\label{eq:gener_propagator2} \ln\hat{g}(z|\kappa) = M  \sum_{\lambda} \rho(\lambda)
\ln\left[1-(1-z)r(\kappa,\lambda) \right].
\end{equation}
The average degree of nodes with hidden variable $\kappa$ is given by the derivative of $\hat{g}(z|\kappa)$ at
$z=1$~\cite{wilf94}, to confirm the obvious
\begin{equation}
\label{eq:avg_k_kappa2} \overline{k}(\kappa) = M \sum_{\lambda} \rho(\lambda) r(\kappa,\lambda),
\end{equation}
while higher moments of $g(k|\kappa)$ can be computed by taking higher order derivatives of the generating function.
%
Eq.~(\ref{eq:avg_k_kappa2}) yields the average degree in the entire top node domain
\begin{equation}
\label{eq:avg_k_2} \overline{k} = \sum_{k} \overline{k}(\kappa) \rho(\kappa) = M \sum_{\kappa,\lambda}
\rho(\kappa)\rho(\lambda) r(\kappa,\lambda),
\end{equation}
and the expected total number of links in the network
\begin{equation}
\label{eq:num_edges} E = N \overline{k} = M \overline{\ell} = N M \sum_{\kappa,\lambda} \rho(\kappa)\rho(\lambda)
r(\kappa,\lambda).
\end{equation}

It is evident from the last equation that to end up with a sparse bipartite network, $E \propto N \propto M$, the connection probability
$r(\kappa,\lambda)$ must be of the form
\begin{equation}
\label{eq:sparse} r(\kappa, \lambda) \propto \hat{r}(\kappa,\lambda) / M,
\end{equation}
where $\hat{r}(\kappa,\lambda)$ is independent of $M$. Therefore, for large sparse networks we can expand the logarithm
in Eq.~(\ref{eq:gener_propagator2}) in powers of $r(\kappa, \lambda)$ to finally obtain, in the first order,
\begin{eqnarray}
{\rm ln}~\hat{g}(z|\kappa) &\approx& (z-1) \sum_\lambda \rho(\lambda) \hat{r}(\kappa,\lambda),\\
\label{propagator2} g(k|\kappa) &=& e^{-\overline{k}(\kappa)}\left[ \overline{k}(\kappa) \right]^{k} /k!,
\end{eqnarray}
which we can use to compute the degree distribution in Eq.~(\ref{eq:pk}).
Propagator $f(\ell|\lambda)$ and degree distribution $P(\ell)$ for bottom nodes
can be obtained from Eqs.~(\ref{propagator2}) and (\ref{eq:pk})
by swapping $\kappa\to\lambda$ and $k\to\ell$.

The Poisson form of the propagator $g(k|\kappa)$, given by Eq.~(\ref{propagator2}), implies that
\begin{equation}
\label{eq:second_moment}
\overline{k^{2}}(\kappa) = \left[\overline{k}(\kappa)\right]^{2} + \overline{k}(\kappa).
\end{equation}
Furthermore, Eq.~(\ref{eq:pk}) allows us to obtain the second moment of the degree distribution:
\begin{equation}
\overline{k^{2}} = \sum_{k} k^{2} P(k) = \sum_{\kappa} \left[\overline{k}(\kappa)\right]^{2} \rho(\kappa) +
\sum_{\kappa}\overline{k}(\kappa)\rho(\kappa)
\end{equation}

\subsection{Unipartite projection}

Next we establish the connection between the degrees of nodes in a bipartite network and in its unipartite projections,
often considered in the literature. In the top unipartite projection, two top nodes are connected if they have at least
one common bottom neighbor in the bipartite network. Therefore, we first compute the probability $p_{0}(\kappa_1,
\kappa_2)$ that two top nodes with hidden variables $\kappa_1$ and $\kappa_2$ do not have any common bottom neighbors
in the bipartite network. This probability is
\begin{equation}
\label{eq:uni_p0} p_{0}(\kappa_1, \kappa_2) = \prod_i \left[1-r(\kappa_1,\lambda_i)r(\kappa_2,\lambda_i)\right],
\end{equation}
where the product is over all the bottom nodes. Taking the logarithm on both sides, we get
\begin{equation}
\ln p_{0}(\kappa_1, \kappa_2) = M \sum_{\lambda} \rho(\lambda)
\ln\left[1-r(\kappa_1,\lambda)r(\kappa_2,\lambda)\right],
\end{equation}
and the probability $p_{u}(\kappa_1, \kappa_2)=1-p_{0}(\kappa_1, \kappa_2)$ that two top nodes with hidden variables $\kappa_1$ and $\kappa_2$ are connected in the
unipartite projection is simply
\begin{equation}
\label{eq:uni_degree_1} p_{u}(\kappa_1, \kappa_2) = 1 - \exp\{ M \sum_\lambda \rho(\lambda)
\ln\left[1-r(\kappa_1,\lambda)r(\kappa_2,\lambda)\right]\}.
\end{equation}
In sparse networks we use Eq.~(\ref{eq:sparse}) to approximate $p_{u}(\kappa_1,\kappa_2)$ as
\begin{equation}
\label{eq:uni_degree_2} p_{u}(\kappa_1, \kappa_2) \approx M \sum_{\lambda}  \rho(\lambda)
r(\kappa_1,\lambda)r(\kappa_2,\lambda).
\end{equation}

Next we find propagator $p(k_u|\kappa)$, the conditional probability that a top node with hidden variable $\kappa$ has
$k_u$ connections in the unipartite projection. The derivation is similar to the derivation of propagator $g(k|\kappa)$
for the bipartite network. We first define partial propagator $p_i^{\kappa'_i}(n_i|\kappa)$, the probability that a top
node with hidden variable $\kappa$ is connected in the unipartite projection to $n_i$ nodes with hidden variable
$\kappa'_i$. Equation~(\ref{eq:uni_degree_2}) indicates that a node with hidden variable $\kappa$ is equally likely to
be connected in the unipartite projection to any of $N_{\kappa'_{i}}$ nodes with hidden variable $\kappa'_i$, where
$N_{\kappa'_i}=N\rho(\kappa'_i)$ is the number of top nodes with hidden variable $\kappa'_i$. If $n_i \ll M$,
we can assume that the links in the unipartite projection are independent, leading to binomial $p_i^{\kappa'_i}(n_i|\kappa)$:
\begin{equation} \label{eq:conditional}
p_i^{\kappa'_i}(n_i|\kappa) = C^{ N_{\kappa'_i}}_{n_i} \left[p_{u}(\kappa,
\kappa'_i)\right]^{n_i}\left[(1-p_{u}(\kappa,\kappa'_i))\right]^{N_{\kappa'_i}-n_i}.
\end{equation}
Similar to Eq.~(\ref{eq:g_propagator}), $p(k_{u}|\kappa)$ is then a convolution
\begin{equation}
\label{eq:uni_propagator} p(k_u|\kappa) = \sum_{\sum n_{i}=k_u} \prod_{i} p_i^{\kappa'_{i}}(n_{i}|\kappa),
\end{equation}
and its generating function $\hat{p}(z|\kappa) = \sum_{k_u}p(k_u|\kappa) z^{k_u}$ is
\begin{equation}
\label{eq:uni_propagator2} \ln\hat{p}(z|\kappa) = N \sum_{\kappa'} \rho(\kappa') \ln\left[ 1-(1-z)
p_{u}(\kappa, \kappa') \right].
\end{equation}
Therefore if $p_{u}(\kappa, \kappa')$ scales as
\begin{equation}
\label{eq:uni_pkk} p_{u}(\kappa, \kappa') \sim {1 \over N^{a}},
\end{equation}
with $a \geq 1$, then similar to the bipartite case, propagator $p(k_u|\kappa)$ is approximately the Poisson
distribution:
\begin{equation}
\label{eq:cond_pk} p(k_u|\kappa) \approx e^{-\overline{k_u}(\kappa)} \left[\overline{k_u}(\kappa)
 \right]^{k_u} /k_u!.
\end{equation}

The average degree $\overline{k_u}(\kappa)$ of nodes with hidden variable $\kappa$ in the unipartite projection
is given by the first derivative of the generating function $\hat{p}(z|\kappa)$ at $z=1$ to yield the obvious
\begin{equation}
\label{eq:avg_uni_k} \overline{k_u}(\kappa) = N \sum_{\kappa'} \rho(\kappa') p_{u}(\kappa,\kappa'),
\end{equation}
which for sparse networks using Eq.~(\ref{eq:uni_degree_2}) transforms to:
\begin{eqnarray}
\label{eq:avg_uni_k2} \overline{k_u}(\kappa) &=& NM \sum_{\lambda,\kappa'} \rho(\lambda) \rho(\kappa') r(\kappa,\lambda) r(\kappa',\lambda)\\
\label{eq:avg_uni_k3}                        &=& M \sum_\lambda \overline{\ell} (\lambda)\rho(\lambda) r(\kappa,\lambda),
\end{eqnarray}
where $\overline{\ell} (\lambda)$ is the average degree of bottom nodes with hidden variable $\lambda$
in the bipartite network.
The average degree in the entire top unipartite projection is then
\begin{equation}
\label{eq:avg_uni_k4} \overline{k_u} = \sum_{\kappa} \rho(\kappa) \overline{k_u}(\kappa) = {M \over N}\sum_{\lambda}
\rho(\lambda) \left[ \overline{\ell}(\lambda)\right]^{2}.
\end{equation}
Finally, the degree distribution in the unipartite projection is
\begin{equation}
\label{eq:uni_pk} P(k_u) = \sum_{\kappa} p(k_u|\kappa) \rho(\kappa).
\end{equation}

\subsection{Number of common neighbors}

The common neighbor statistics is useful in many applications, such as node similarity estimation~\cite{leight06}  and
link prediction~\cite{adamic03}.
We compute the probability that two top nodes with hidden variables $\kappa_1$ and $\kappa_2$ have $m$ common bottom
neighbors. This probability can be calculated as
\begin{equation}
\label{eq:pm_propagator} P_{\kappa_1, \kappa_2}(m) = \sum_{\sum m_{i}=m} \prod_{i}
p_{\kappa_1,\kappa_2}(m_{i}|\lambda_{i}),
\end{equation}
where $p_{\kappa_1,\kappa_2}(m_{i}|\lambda_{i})$ is the probability that two top nodes with $\kappa_1$ and $\kappa_2$
have $m_i$ common bottom neighbors with $\lambda_i$, and the product is over the entire range of $\lambda_i$, while the
summation is over all possible combinations of $m_{i}$ adding up to $m$.

Consider two nodes with hidden variables
$\kappa_1$ and $\kappa_2$. Each common neighbor of the two nodes with $\kappa_1$ and $\kappa_2$ appears independently
with probability
\begin{equation}
\tilde{r}_{\lambda}(\kappa_1,\kappa_2)= r(\kappa_1,\lambda)r(\kappa_2,\lambda),
\end{equation}
where $\lambda$ is the hidden variable of the common neighbor. Therefore, $p_{\kappa_1,\kappa_2}(m|\lambda)$ is also
binomial:
\begin{equation}
p_{\kappa_1,\kappa_2}(m|\lambda) = C^{M_{\lambda}}_{m} \left[\tilde{r}_{\lambda}(\kappa_1,\kappa_2)) \right]^{m}
\left[1-\tilde{r}_{\lambda}(\kappa_1,\kappa_2) \right]^{M_{\lambda}-m},
\end{equation}
and the corresponding generating function is given by
\begin{equation}
\widehat{p}_{\kappa_1,\kappa_2}(z|\lambda) = \left[1-(1-z)\tilde{r}_{\lambda}(\kappa_1,\kappa_2) \right]^{M_\lambda}.
\label{eq:avg_m_binom_gener}
\end{equation}
Since $P_{\kappa_1, \kappa_2}(m)$ is given by a convolution, its generation function is
\begin{equation}
\label{eq:p_generat_product} \widehat{P}_{\kappa_1,\kappa_2}(z) =
\prod_{i}\widehat{p}_{\kappa_1,\kappa_2}(z|\lambda_{i}).
\end{equation}
Combining the last two equations, we get
\begin{equation}
\label{eq:p_generat2} \ln\widehat{P}_{\kappa_1,\kappa_2}(z)= M \sum_\lambda \rho(\lambda)
\ln\left[1-(1-z)\tilde{r}_{\lambda}(\kappa_1,\kappa_2) \right].
\end{equation}

To compute the average number of common neighbors between top nodes with $\kappa_1$ and $\kappa_2$ we evaluate the
derivative of $\widehat{P}_{\kappa_1,\kappa_2}(z)$ with respect to $z$ at $z=1$:
\begin{equation}
\label{eq:avg_m} \overline{m}(\kappa_1,\kappa_2) = M \sum_\lambda \rho(\lambda) \tilde{r}_{\lambda}(\kappa_1,\kappa_2).
\end{equation}
The generating function for the common neighbor distribution has the same structure as $\hat{g}(z|k)$. Therefore, the
closed form of $P_{\kappa_1,\kappa_2}(m)$ in the sparse network approximation is given by
\begin{equation}
\label{eq:p_generat4} P_{\kappa_1, \kappa_2}(m) \approx e^{-\overline{m}(\kappa_1, \kappa_2)} \left[
\overline{m}(\kappa_{1},\kappa_{2}) \right]^{m}/m!.
\end{equation}

\subsection{Degree correlations}

The degree correlations in bipartite networks are fully described by conditional probabilities $P(\ell|k)$ and
$P(k|\ell)$ in Eqs.~(\ref{eq:cond_prob1},\ref{eq:cond_prob2}). In order to calculate $P(\ell|k)$ we need to
define the related conditional probability $\rho(\lambda|\kappa)$ that an edge outgoing from a top node with hidden
variable $\kappa$ is connected to a bottom node with hidden variable $\lambda$. Then, $P(\ell|k)$ can be written as
\begin{equation}
\label{eq:corr_pk} P(\ell|k) = \sum_{\kappa,\lambda} f(\ell-1|\lambda) \rho(\lambda|\kappa) g^{*}(\kappa|k),
\end{equation}
where $f(\ell-1|\lambda)$ is the conditional probability that a bottom node with hidden variable $\lambda$ ends up
having degree $\ell$ (one connection is already taken into account by the conditional edge), while the inverse
propagator $g^{*}(k|\kappa)$ is the probability that a top node of degree $k$ has hidden variable $\kappa$. This
inverse propagator is given by the Bayes' formula~\cite{gnedenko62}
\begin{equation}
\label{eq:bayes} P(k)g^{*}(\kappa|k) = \rho(\kappa) g(k|\kappa),
 \end{equation}
using which we write
\begin{equation}
\label{eq:corr_pk2} P(\ell|k) = {1 \over P(k)} \sum_{\kappa,\lambda} \rho(\kappa) \rho(\lambda|\kappa)
f(\ell-1|\lambda) g(k|\kappa).
\end{equation}
To determine $\rho(\lambda|\kappa)$ we note that the conditional probability that an edge is connected to a bottom node
with $\lambda$, given that this edge is connected to a top node with $\kappa$, is proportional to the density of bottom
nodes $\rho(\lambda)$ and the connection probability $r(\kappa,\lambda)$,
\begin{equation}
\label{eq:cond_corr} \rho(\lambda|\kappa) \propto \rho(\lambda) r(\kappa, \lambda).
\end{equation}
Taking into account the normalization condition $\sum_{\lambda} \rho(\lambda|\kappa) = 1$, we get
\begin{equation}
\label{eq:cond_corr2} \rho(\lambda|\kappa) = { \rho(\lambda) r(\kappa, \lambda) \over \sum_{\lambda'} \rho(\lambda')
r(\kappa, \lambda')}.
\end{equation}
Using Eqs.~(\ref{eq:corr_pk2}-\ref{eq:cond_corr2}) we obtain the final expression
for the top ANND statistics:
\begin{equation}
\label{eq:ANND} \overline{\ell}_{nn}(k) = \sum_{\ell} \ell P(\ell|k)= 1 + {1\over P(k)}\sum_\kappa
\overline{\ell}_{nn}(\kappa) \rho(\kappa) g(k|\kappa),
\end{equation}
where $\overline{\ell}_{nn}(\kappa)$ is the average nearest neighbor degree of top nodes with hidden variable $\kappa$:
\begin{equation}
\label{eq:ANND_kappa} \overline{\ell}_{nn}(\kappa) = \sum_{\lambda} \overline{\ell}(\lambda) \rho(\lambda|\kappa).
\end{equation}

\subsection{Bipartite clustering coefficient}
Finally we derive the bipartite clustering coefficient as defined by P. Zhang et al~\cite{zhang08}. Other
variations of the bipartite clustering coefficient can be computed in a similar manner.

The bipartite clustering coefficient of top node $i$, given by Eq.~(\ref{eq:clustering}), can be written as
\begin{equation}
\label{eq:cBi} c_{B}(i) =  {\sum_{j > l}(m_{jl}-1)  \over \sum_{j > l} (k_{j}+k_{l} - m_{jl} - 1)},
\end{equation}
where $m_{jl}$ is the number of common neighbors between bottom nodes $j$ and $l$, while $k_j$ and $k_l$ are their degrees.
Since the summations in the
numerator and denominator are performed independently, we can estimate the
average bipartite clustering coefficient of top nodes with hidden variable $\kappa$
by calculating the ensemble averages of the numerator and denominator.
The details are in the Appendix, while the answer is
\begin{equation}
\label{eq:cBi_final} \overline{c_{B}}(\kappa) = {
\sum_{\lambda_1,\lambda_2}\rho(\lambda_{1}|\kappa)\rho(\lambda_{2}|\kappa) \overline{m}(\lambda_1,\lambda_2) \over 2
\overline{\ell}_{nn}(\kappa) - \sum_{\lambda_1,\lambda_2}\rho(\lambda_{1}|\kappa)\rho(\lambda_{2}|\kappa)
\overline{m}(\lambda_1,\lambda_2)},
\end{equation}
where $\rho(\lambda|\kappa)$ is the conditional probability that an edge connected to a top node with hidden variable
$\kappa$ is also connected to a bottom node with hidden variable $\lambda$, $\overline{m}(\lambda_1, \lambda_2)$ is the
average number of common neighbors between two bottom nodes with hidden variables $\lambda_1$ and $\lambda_2$,
and $\overline{\ell}_{nn}(\kappa)$ is the average nearest neighbor degree of top nodes with hidden variable $\kappa$.
The average bipartite clustering coefficient of top nodes with degrees $k \geq 2 $ can be expressed in terms of
$\overline{c_B}(\kappa)$ as
\begin{equation}
\label{eq:avg_cBi_k} \overline{c_{B}}(k) = {1 \over P(k)} \sum_\kappa \rho(\kappa) g(k|\kappa)
\overline{c_{B}}(\kappa),
\end{equation}
while the average bipartite clustering coefficient in the
top node domain is simply
\begin{equation}
\label{eq:cB_total} \overline{c_B} = \sum_{\kappa} \rho(\kappa) \overline{c_B}(\kappa) = \sum_k P(k) \overline{c_{B}}(k).
\end{equation}

\section{Examples of Bipartite Networks with Hidden Variables}

Having the general formalism in place, we next consider a couple of examples of bipartite networks with hidden variables.
The first example of uncorrelated networks is fairly standard. The second one, stratified networks, is more unusual.

\subsection{Uncorrelated Bipartite Networks}

Consider a random bipartite network composed of nodes with degrees $\{k_{i}\}$ and $\{\ell_{j}\}$ drawn from distributions $P(k)$
and $P(\ell)$. If nodes in the network are connected at random, then two randomly chosen nodes with degrees $k$ and $\ell$
are connected with probability $p = {k \ell /E}$, where $E$ is the total number of links in the network.

Similar random uncorrelated networks can be constructed in the hidden variable formalism. Consider a network
with hidden variables drawn from distributions $\rho(\kappa)$ and $\rho(\lambda)$, in which node pairs are connected with
probability proportional to the product of nodes' hidden variables:
\begin{equation}
\label{eq:random_r} r(\kappa, \lambda) = {\kappa \lambda \over C},
\end{equation}
where $C$ is some normalization constant.
The above form of $r(\kappa, \lambda)$ implies that the hidden variable of a node can be regarded as its target or expected degree.
Indeed, if we choose $C=\overline{\lambda}M$, then
a top node with hidden variable $\kappa$ gets $\kappa$ connections on average
\begin{equation}
\label{eq:random_k_kappa} \overline{k}(\kappa) = M \sum_\lambda \rho(\lambda) r(\kappa, \lambda) = \kappa.
\end{equation}

Since the assumption of a sparse network, given by Eq.~(\ref{eq:sparse}) holds here, propagator $g(k|\kappa)$ is given
by the Poisson distribution:
\begin{equation}
g(k|\kappa) = e^{-\kappa} \kappa^{k}/k!,
\end{equation}
and using Eqs.~(\ref{eq:pk}) and (\ref{eq:second_moment}) one can obtain
\begin{equation}
\overline{k^{2}} = \overline{\kappa^{2}} + \overline{\kappa}.
\end{equation}

 One type of nodes in real bipartite networks is often characterized by scale-free degree distributions, while
degree of nodes of the other type can follow either fat-tailed or poissonian distributions~\cite{guillaume04}. Our
uncorrelated formalism can account for both options. The former case is actually simpler, and the properties of top and
bottom nodes can be obtained from each other via a simple swap of notations. Therefore below we consider the latter
case, which is more typical for real networks.

Specifically, let $\kappa$ be power-law distributed:
\begin{equation}
\label{eq:random_rho_kappa} \rho(\kappa) = (\gamma-1) k_{0}^{\gamma-1} \kappa ^{-\gamma},
\end{equation}
where power-law exponent $\gamma$ and minimum expected degree $\kappa_0$ are parameters of the distribution. The
resulting degree distribution of the top node domain is given by Eqs.~(\ref{eq:pk}) and~(\ref{propagator2}), which
yield
\begin{equation}
\label{eq:random_pk} P(k) = (\gamma - 1) \kappa_{0}^{\gamma-1} {\Gamma[k-\gamma + 1,\kappa_0] \over \Gamma[k+1]},
\end{equation}
where $\Gamma[x,s]$ is the incomplete gamma function. In the large $k$ limit we can approximate $P(k)$ by
\begin{equation}
\label{eq:random_pk2} P(k) \approx (\gamma - 1) \kappa_{0}^{\gamma-1} k^{-\gamma}.
\end{equation}
We note that the distribution $P(k)$ of top node degrees does not depend on a specific form
of the hidden variable distribution $\rho(\lambda)$ in the bottom node domain.
Let the latter be a delta function $\rho(\lambda) = \delta(\lambda-\lambda_{0})$, meaning that
all bottom nodes have the same value of their hidden variable equal to $\lambda_{0}$. Then using
the same Eqs.~(\ref{eq:pk},\ref{propagator2}) swapped for the bottom nodes, we immediately conclude
that the distribution of bottom node degrees is poissonian:
\begin{equation}
\label{eq:random_pl} P(\ell) = e^{-\lambda_0} \lambda_{0}^{\ell} / \ell!.
\end{equation}

We now turn our attention to the unipartite projections. We first consider the top node
projection. We use Eq.~(\ref{eq:avg_uni_k3}) to compute the average degree of $\kappa$-nodes:
\begin{equation}
\label{eq:random_uni_k} \overline{k}_{u}(\kappa) = \kappa \lambda_0.
\end{equation}
Therefore the average degree in the top unipartite projection is
\begin{equation}
\label{eq:random_uni_avg_k} \overline{k}_{u} = \overline{\kappa} \lambda_0.
\end{equation}
The degree distribution in the projection are given by Eqs.~(\ref{eq:uni_pk}) and~(\ref{eq:cond_pk}):
\begin{equation}
\label{eq:random_uni_pk} P(k_u) = (\gamma-1) [\kappa_0 \lambda_0]^{\gamma-1} {\Gamma[k_u -\gamma + 1, \kappa_0
\lambda_0]\over \Gamma[k_{u}+1]},
\end{equation}
that is, this distribution is also a power law,
\begin{equation}
P(k_u) \sim (\gamma-1) [\kappa_0 \lambda_0]^{\gamma-1} k_u^{-\gamma},
\end{equation}
and the exponent of this power law is equal to the exponent of the top power-law degree distribution in the original
bipartite network.

In the bottom unipartite projection, the average node degrees are obtained in a similar manner to yield
\begin{equation}
\label{eq:random_uni2_avg_k} \overline{\ell_{u}} = \overline{\ell_{u}}(\lambda) = \lambda_0 {\overline{\kappa^{2}}
\over \overline{\kappa}}.
\end{equation}
For $\gamma \leq 3$, $\overline{\kappa^{2}}$ depends on $N$, and diverges in the thermodynamic limit.
Therefore connection probability $p(\lambda_1,\lambda_2)$ does not satisfy the condition of
Eq.~(\ref{eq:uni_pkk}), and we can not approximate the degree distribution in the bottom domain by
Eq.~(\ref{eq:uni_pk}) with poissonian $p(k_u|\kappa)$ in Eq.~(\ref{eq:cond_pk}). However, if $\gamma > 3$,
then $\overline{\kappa^{2}}$ is finite in the thermodynamic limit, and the degree distribution
is given by
\begin{equation}
\label{eq:random_uni2_pk} P(\ell_{u}) = e^{- \overline{\ell_{u}}} [ \overline{\ell_{u}}]^{\ell} /\ell!.
\end{equation}

As far as correlations are concerned, the conditional hidden variable distributions are
\begin{eqnarray}
\label{eq:random_rho_lambda} \rho(\lambda|\kappa) &=& \delta(\lambda-\lambda_0),\\
\label{eq:random_pll} \rho(\kappa|\lambda) &=& {\kappa \over \overline{\kappa}} \rho(\kappa),
\end{eqnarray}
leading to the following expression for the top and bottom ANNDs
given by Eq.~(\ref{eq:ANND}):
\begin{eqnarray}
\label{eq:random_k_annd} \overline{\ell}_{nn}(k) &=& 1+\lambda_0,\\
\label{eq:random_l_annd} \overline{k}_{nn}(\ell) &=& 1+{{\overline{\kappa^{2}}} \over \overline{\kappa}} =
{\overline{k^{2}} \over \overline{k}}.
\end{eqnarray}

The average number of common neighbors is given by Eq.~(\ref{eq:avg_m}) yielding,
for top and bottom nodes,
\begin{eqnarray}
\label{eq:random_avg_m1} \overline{m}(\kappa_{1},\kappa_{2}) &=& {\kappa_{1} \kappa_{2} \over M},\\
\label{eq:random_avg_m2} \overline{m}(\lambda_0,\lambda_0)   &=& {\lambda_{0}^{2} \over N} {\overline{\kappa^{2}} \over
\overline{\kappa}^{2}}.
\end{eqnarray}

Finally, to compute clustering, we insert the expressions for the average number of common neighbors
(\ref{eq:random_avg_m1},\ref{eq:random_avg_m2}), ANNDs~(\ref{eq:random_k_annd},\ref{eq:random_l_annd}), and conditional
distributions (\ref{eq:random_rho_lambda},\ref{eq:random_pll}) into Eq.~(\ref{eq:cBi_final}), and obtain the average
bipartite clustering coefficient for top and bottom nodes:
\begin{eqnarray}
\label{eq:random_ck} \overline{c}_{B}(\kappa) &=& {{(\lambda_{0}^{2} \overline{\kappa^{2}}) / (N \overline{\kappa}^{2})}
\over 2\lambda_0 - {(\lambda_{0}^{2} \overline{\kappa^{2}}) / (N \overline{\kappa}^{2})}}
\approx {\lambda_{0} \over 2 N} {\overline{\kappa^{2}} \over \overline{\kappa}^{2}},\\
\overline{c}_{B}(\lambda) &=& {{ (\overline{\kappa^{2}})^{2} / (M \overline{\kappa}^{2})} \over {{2\overline{\kappa^{2}}}
/ \overline{\kappa}} - { (\overline{\kappa^{2}})^{2} / (M \overline{\kappa}^{2})}}\approx {\overline{\kappa^{2}}
\over 2 M \overline{\kappa}}.
\end{eqnarray}
We observe that the clustering coefficient of a node does not depend on its hidden variable in either case, i.e., that
it is constant. This constant decreases as the network sizes $N,M$ increase, and vanishes in the thermodynamic
limit.

\begin{figure}
\includegraphics[width=8.0 cm,angle=0]{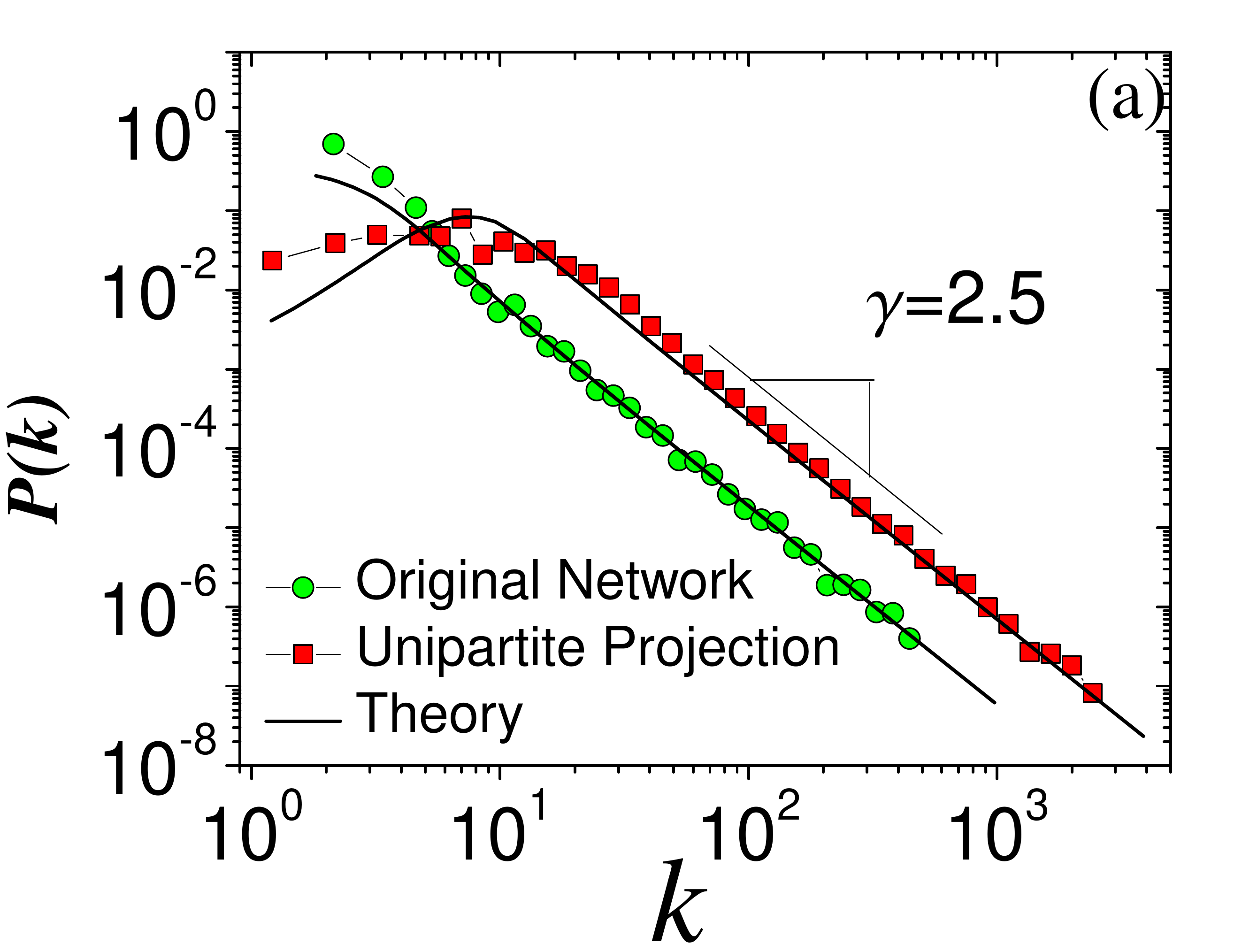}
\includegraphics[width=8.0 cm,angle=0]{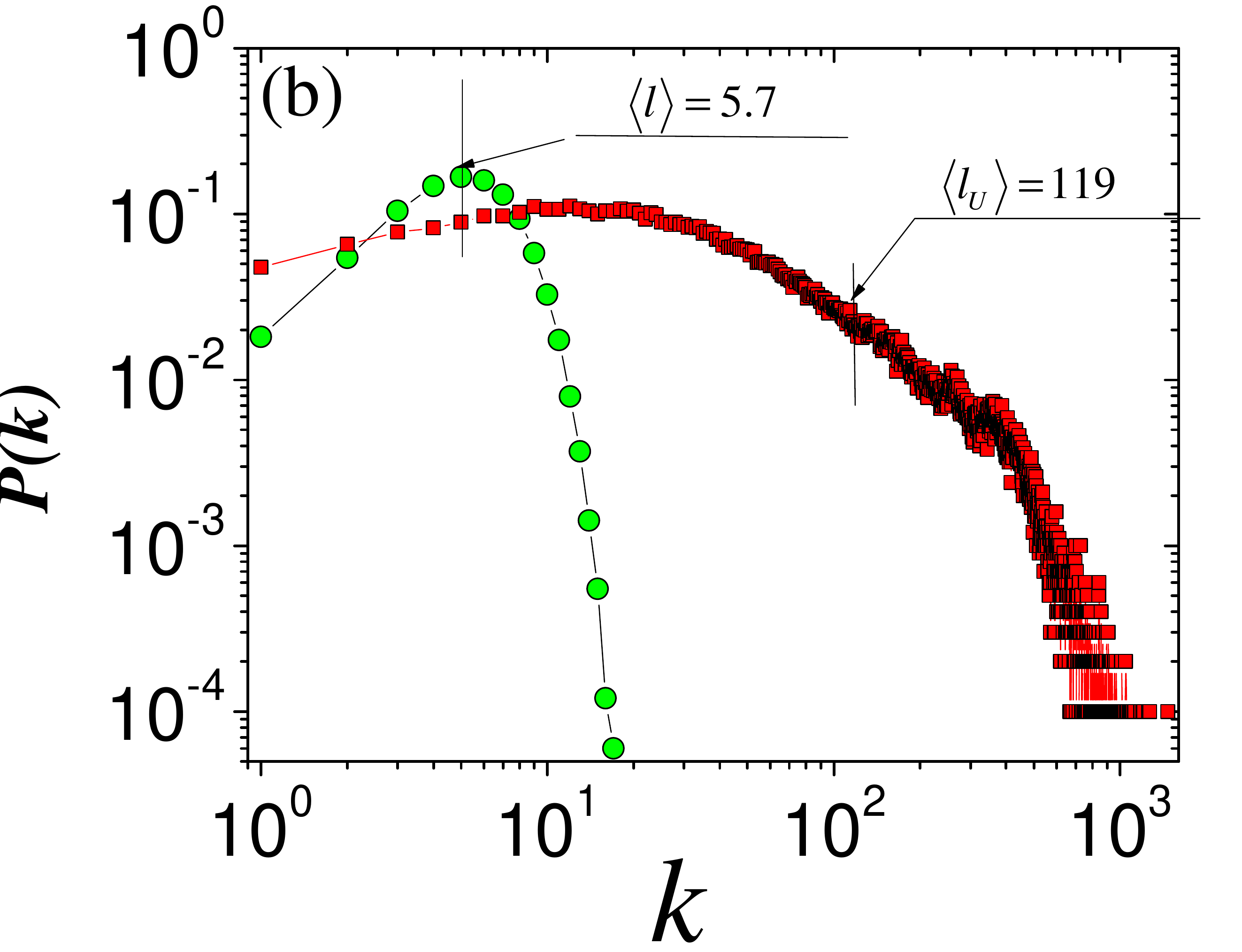}
\caption{ \footnotesize (Color Online) \label{fig:unc1} Degree distributions in a random uncorrelated bipartite
network. (a)~Degree distributions in the top domain (green circles) and top unipartite projection (red squares). The
solid lines are the analytical predictions from Eqs.~(\ref{eq:random_pk},\ref{eq:random_uni_pk}). (b)~Degree
distributions in the bottom domain (green circles) and bottom unipartite projection (red squares). Both plots
correspond to the model with $N=2 \times 10^{5}$, $M = 10^{5}$, $\gamma=2.5$, $\kappa_0=1$, and $\lambda_0 = 6$.}
\end{figure}
To test our analytical results we perform simulations, setting $N=2M$, $N=2\times10^{5}$, $\gamma=2.5$, $\kappa_{0} =
1$, and $\lambda_0 = 6$ to satisfy $N \overline{k} = M \overline{\ell}$. The degree distributions in the top and bottom
domains as well as in their unipartite projections are shown in Fig.~(\ref{fig:unc1}). The degree distribution of top
nodes in the original bipartite network, and in its top unipartite projection both follow a power law with the same
exponent $\gamma=2.5$, see Fig.~\ref{fig:unc1}(a). As seen in Fig.~\ref{fig:unc1}(b), the degree distribution in the
bottom node domain is well approximated by a Poisson distribution. On the other hand, due to the divergent behavior of
the second moment of the top degree distribution $\overline{\kappa^{2}}$, the degree distribution in the bottom
unipartite projection seems to follow a truncated power-law. The measured values of $\overline{k_{u}} = 20.0$ and
$\overline{\ell_{u}} \approx 119$ are in good agreement with
Eqs.~(\ref{eq:random_uni_avg_k},\ref{eq:random_uni2_avg_k}) since $\overline{\kappa}=2.85$ and $\overline{\kappa^2}=62$
for the selected parameters.

\begin{figure}
\includegraphics[width=8.0 cm,angle=0]{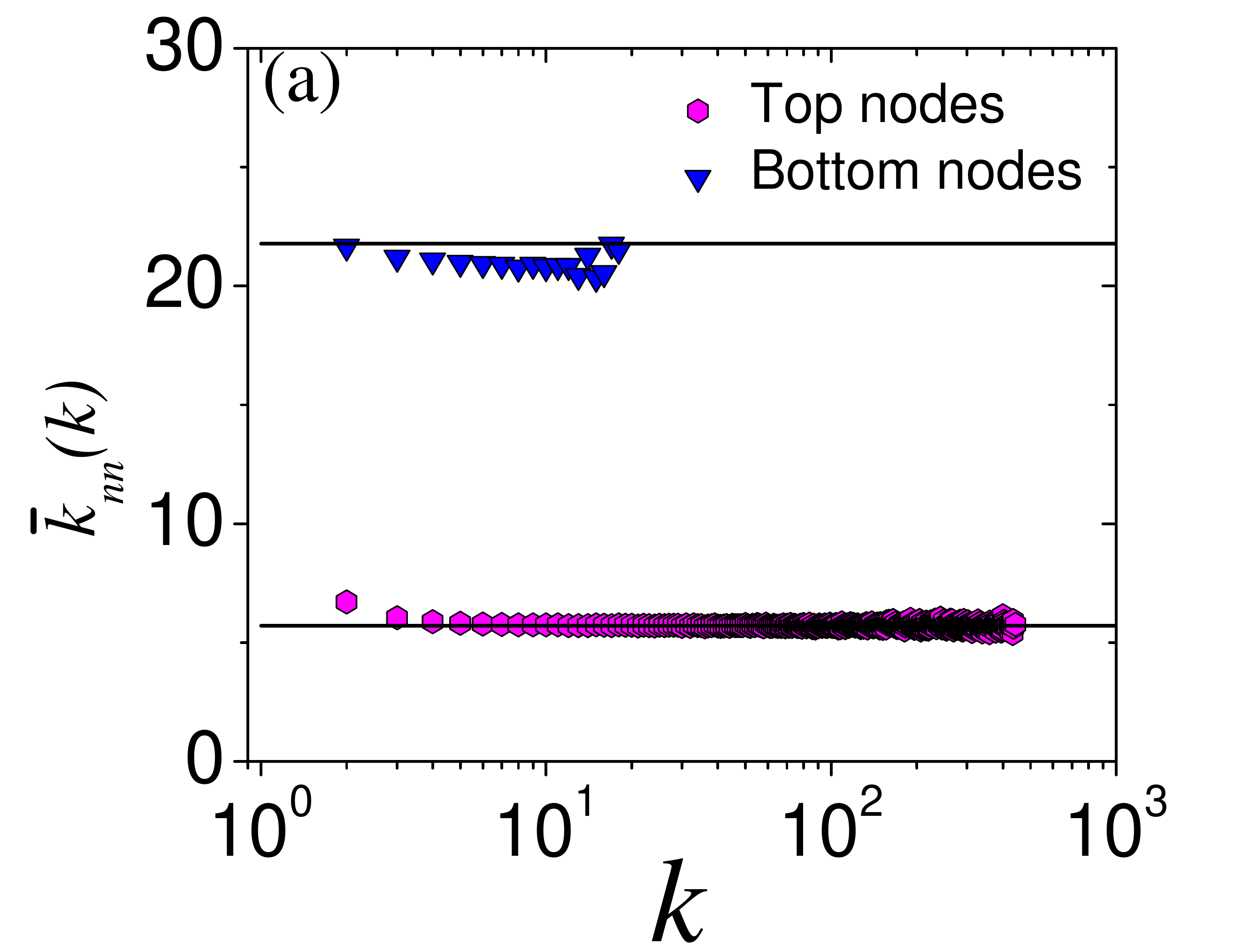}
\includegraphics[width=8.0 cm,angle=0]{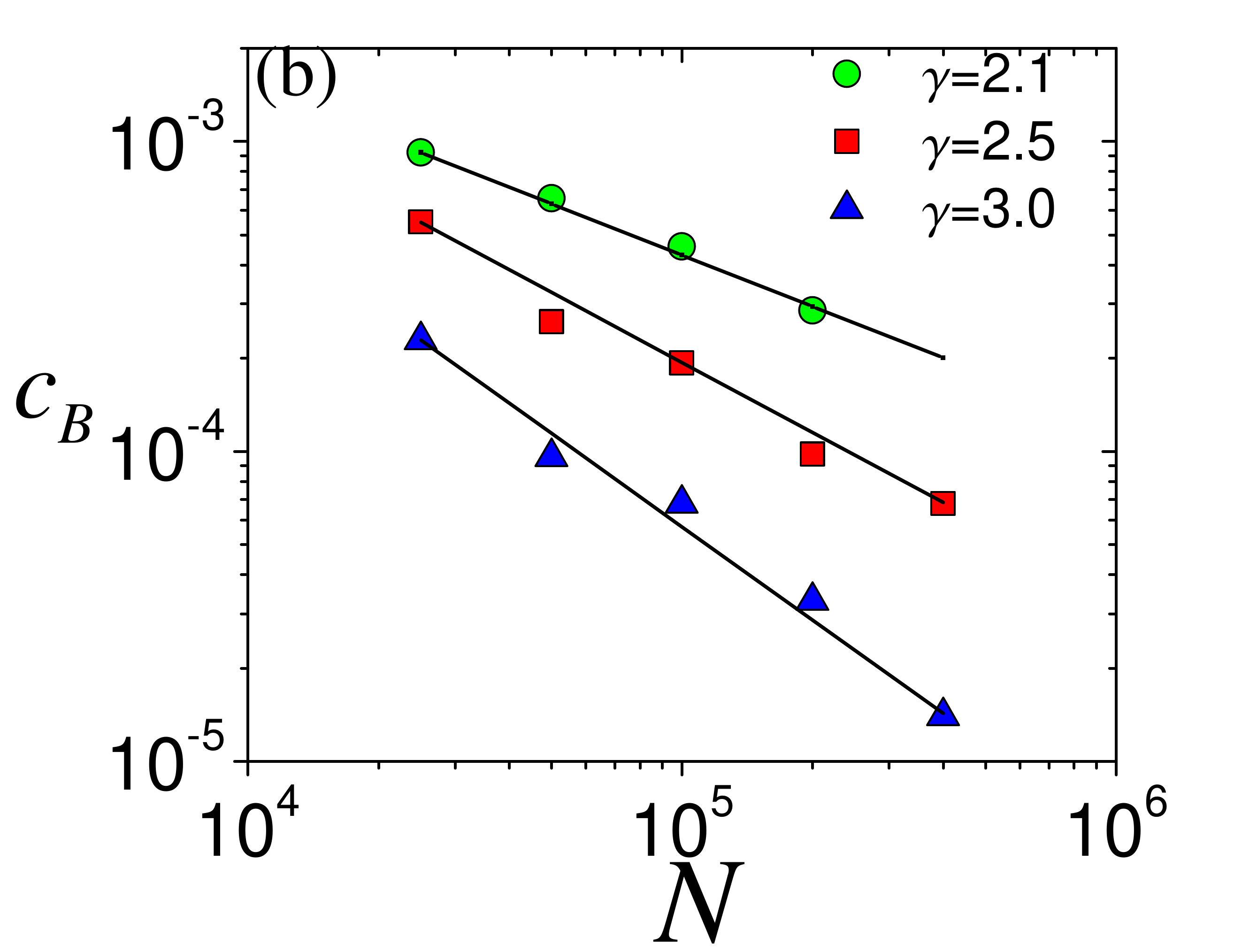}
\caption{ \footnotesize (Color Online) \label{fig:unc2}(a) The average nearest neighbor degrees of top (blue triangles)
and bottom (magenta hexagons) nodes in an uncorrelated bipartite network with $N=2\times10^{5}$, $M = 10^{5}$,
$\gamma=2.5$, $\kappa_0=1$, and $\lambda_0=6.0$. The solid lines are the analytical predictions in
Eqs.~(\ref{eq:random_k_annd},\ref{eq:random_l_annd}). (b) The average bipartite clustering coefficient for top nodes in
uncorrelated bipartite networks as a function of network size $N$ for $\gamma=2.1$, $\gamma=2.5$ and $\gamma = 3.0$.
The solid lines are the theoretical predictions of $\overline{c}_{B}\sim N^{-\delta}$ with $\delta = (\gamma-1)/2$.}
\end{figure}
In Fig.~\ref{fig:unc2}(a) we plot the ANNDs, and confirm that they are independent of node degrees as
Eqs.~(\ref{eq:random_k_annd},\ref{eq:random_l_annd}) predict for uncorrelated networks.

To test the dependence of the average bipartite clustering coefficient on the network size, we generate a number of
uncorrelated bipartite network of different sizes and values of $\gamma$. While sampling hidden variables $\kappa$ for
top nodes, we impose the cutoff of $\kappa_{max} \sim N^{1/2}$ to avoid structural degree correlations~\cite{burda03}.
Therefore, $\overline{\kappa^{2}}\sim N^{(3-\gamma)/2}$, and the average bipartite clustering coefficient scales as
$\overline{c}_{B} \sim N^{-\delta}$ with $\delta=(\gamma-1)/2$ for $2<\gamma<3$. In Fig.~\ref{fig:unc2}(b) we confirm
this scaling. The figure shows the measured bipartite clustering coefficients as a function of $N$ for different values
of $\gamma$.

\subsection{Stratified Bipartite Networks}

The original stratified unipartite network model was considered by Leicht et al~\cite{leight06}. In this model, $N$
nodes are assigned random integer {\em ages} $t_{i}= 1,\ldots,t_{max}$
with uniform probability, and then links are created between node pairs with probability
\begin{equation}
\label{eq:strat_conn} P(\Delta t) = p_{0} e^{-a|\Delta t|},
\end{equation}
where $p_0$ and $a$ are model parameters. The motivation for this model in~\cite{leight06} was
to have a simplified social model in which
individuals preferably connect to other individuals of similar age. The stratified model was used in~\cite{leight06}
to test the ability of different node similarity measures to infer relative node ages.

Here we generalize the stratified network model as follows. The networks in the model consist of $N$ top and $M$ bottom
nodes. All nodes are assigned hidden variables $\kappa$ and $\lambda$ drawn from the continuous uniform distribution on
interval $[0,T]$, $\rho(\kappa) = \rho(\lambda) = {1/T}$. To eliminate finite size effects we impose the periodic
boundary condition, meaning that nodes are uniformly scattered along a circle, and their hidden variables are their
angular coordinates if we set $T=2\pi$. To simplify the calculations we use the squared distances in the connection probability function:
\begin{equation}
\label{eq:strat_conn2} r(\kappa, \lambda) = r_{0} e^{-a \|\lambda-\kappa\|^{2}},
\end{equation}
where $\|\lambda - \kappa \|$ is the angular distance between $\lambda$ and $\kappa$:
\begin{equation}
\|\lambda - \kappa \| = \pi - |\pi - |\lambda - \kappa||.
\end{equation}

We first calculate the degree distributions for the top nodes. Due to the uniform distribution of hidden
variables, the expected degree of a node is independent of its hidden variable $\kappa$. Using
Eqs.~(\ref{eq:avg_k_kappa2}) and (\ref{eq:avg_k_2}) we obtain
\begin{equation}
\label{eq:strat_k} \overline{k} = \overline{k}(\kappa) = {M r_{0} \over 2 \sqrt{\pi a}} {\rm Erf}(\pi \sqrt{a}),
\end{equation}
where ${\rm Erf}(x)$ is the error function. For $\overline{k}$ to be independent of network size, we must set
${r_0 / \sqrt{a}} \sim 1/M$. Another natural choice would be to constraint $r_{0} = M^{-1}$, but this
choice would lead to bipartite clustering coefficients dependent on the network size.
Constant bipartite clustering can be instrumented by setting
\begin{equation}
\label{eq:strat_param} r_{0}=1, \quad\text{and}\quad a = \widetilde{a} M^{2},
\end{equation}
where $\widetilde{a}$ is a parameter controlling the average degree in the network. With the above choice of
parameters  Eq.~(\ref{eq:strat_k}) simplifies to
\begin{equation}
\label{eq:strat_k2} \overline{k} = \overline{k}(\kappa) \approx {1 \over 2 \sqrt{ \pi\widetilde{ a}}}.
\end{equation}
Similarly, the average degree in the bottom node domain is given by
\begin{equation}
\label{eq:strat_k_bottom} \overline{\ell} = \overline{\ell}(\lambda) = {N \over M}\overline{k}.
\end{equation}

Since connection probability $r(\kappa,\lambda)$ does not scale as $M^{-1}$,
propagator $g(k|\kappa)$ is not given by Eq.~(\ref{propagator2}). Instead we
have to use Eq.~(\ref{eq:gener_propagator2}) to compute the propagator, yielding
\begin{equation}
\label{eq:strat_gener} \hat{g}(z|\kappa) = e^{- \overline{k} {\rm Li}_{3/2}(1-z)},
\end{equation}
where ${\rm Li}_{n}(x)$ is the polylogarithm. Equation~(\ref{eq:strat_gener}) can be used to calculate higher moments
of the degree distribution. For example, the second moment is
\begin{equation}
\label{eq:avg_kk} \overline{k^{2}} = \overline{k}^{2} + \overline{k}(1-{1 \over \sqrt{2}}).
\end{equation}
That is, similar to the Poisson distribution, the standard deviation of $g(k|\kappa)$ is
\begin{equation}
\label{eq:sigma_k} \sigma = \sqrt{\overline{k^{2}} - \overline{k}^{2}} \propto
\sqrt{\overline{k}}.
\end{equation}

According to Eq.~(\ref{eq:ANND_kappa}), the average nearest neighbor degree is independent of the node's hidden variable:
\begin{equation}
\label{eq:strat_lnn} \overline{\ell}_{nn}(\kappa) = \overline{\ell},
\end{equation}
because node degrees are not correlated with their hidden variables, see Eq.~(\ref{eq:strat_k2}).  Therefore,
despite strong correlation between hidden variables of connected nodes, there are no degree correlations. The ANND can
be obtained by inserting $\overline{\ell}_{nn}(\kappa)$ from Eq.~(\ref{eq:strat_lnn}) into Eq.~(\ref{eq:ANND}) to yield
\begin{equation}
\overline{\ell}_{nn}(k) = 1+\overline{\ell}.
\end{equation}

\begin{figure}[!ht]
\includegraphics[width=8.0 cm,angle=0]{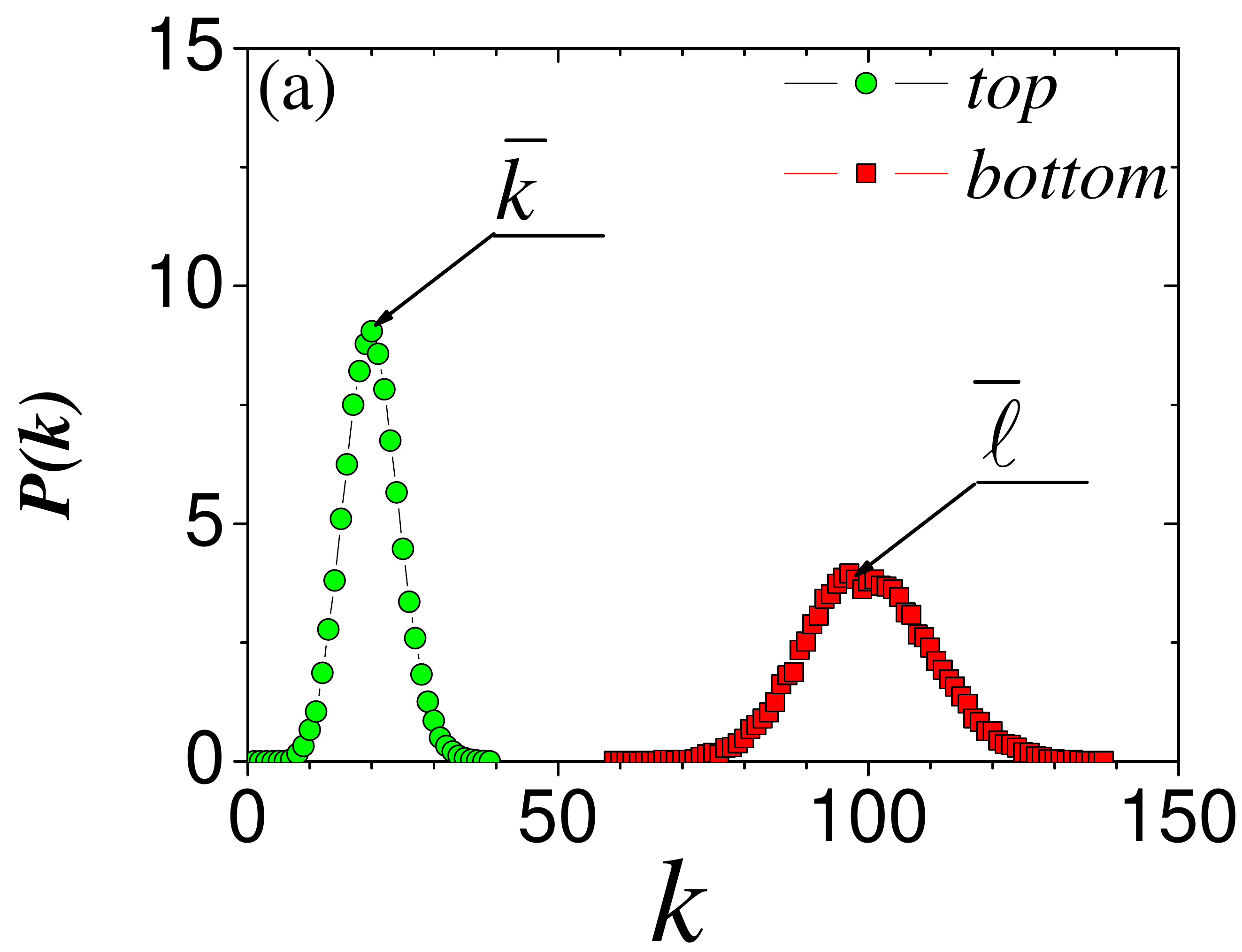}
\includegraphics[width=8.0 cm,angle=0]{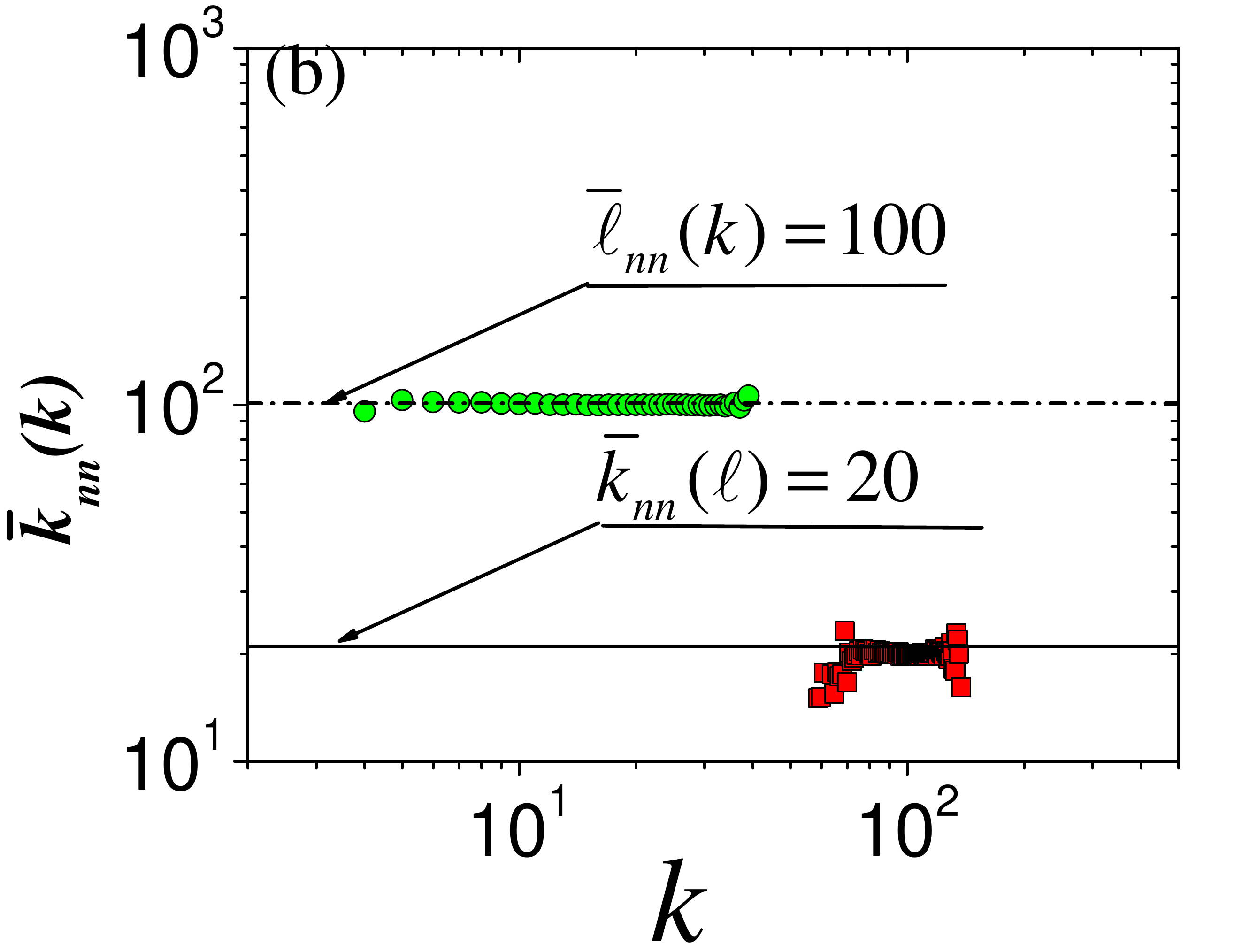}
\includegraphics[width=8.0 cm,angle=0]{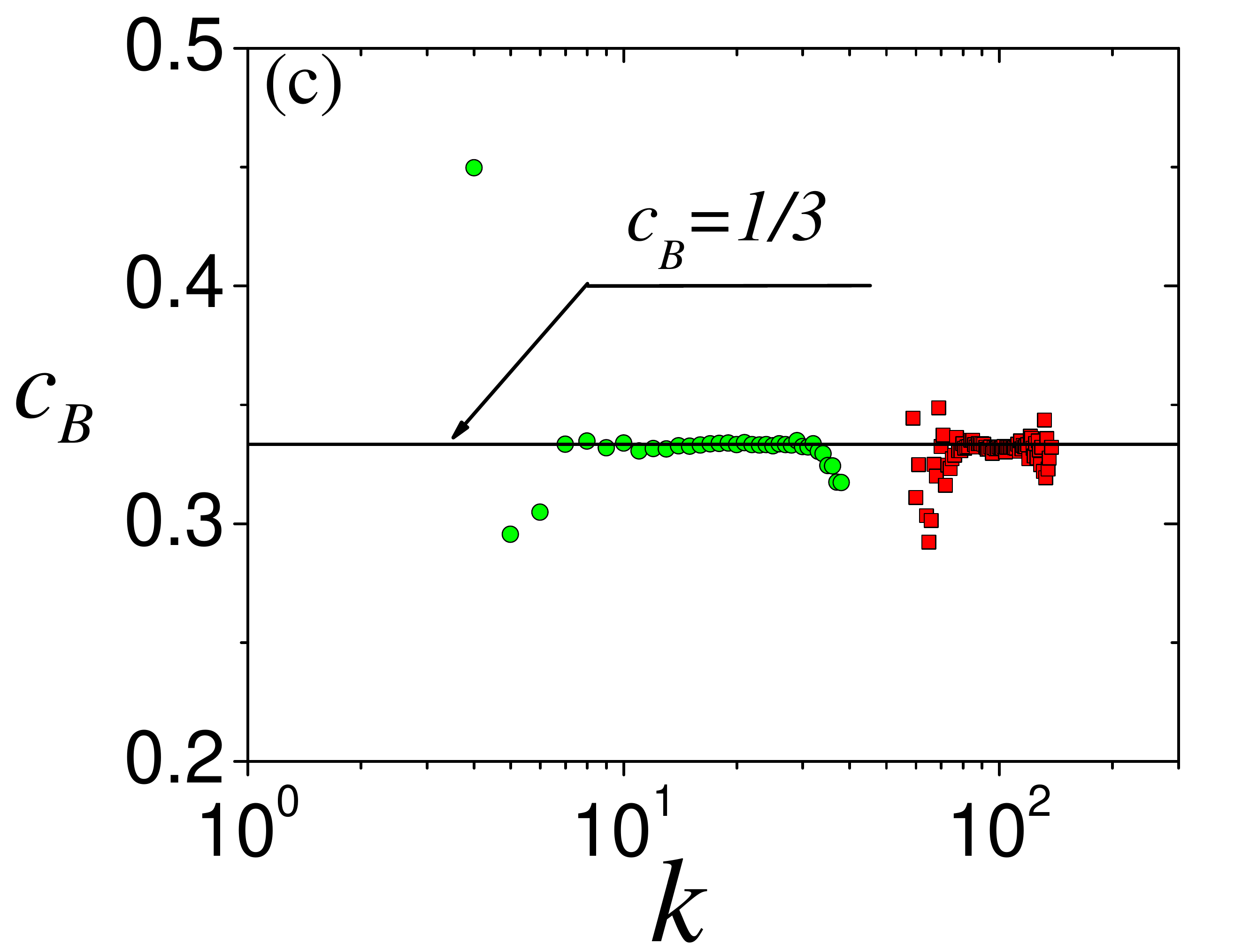}
\caption{ \footnotesize (Color Online) Stratified bipartite networks. (a) Degree distributions of top (green circles)
and bottom (red rectangles) nodes. (b) Average nearest neighbor degrees for top and bottom nodes as a function of node
degree. (c) Average bipartite clustering coefficients of top and bottom nodes as a function of node degree. All the
plots are for stratified bipartite networks with $N=10^{5}$, $M=2\times10^{5}$, and $\overline{k} = 20$.
\label{fig:strat}}
\end{figure}

The average number of common neighbors between bottom nodes with hidden variables $\lambda_1$ and $\lambda_2$
is given by Eq.~(\ref{eq:avg_m}), which now becomes
\begin{equation}
\label{eq:strat_avg_m} \overline{m}(\lambda_1, \lambda_2) = {N p_{0}^{2} \over 2\pi} \int_{-\pi}^{\pi} e^{-a \|
\lambda_1 - \kappa \|^{2} }e^{-a\|\lambda_2 - \kappa \|^{2}} {\rm d} \kappa.
\end{equation}
To compute $\overline{m}(\lambda_1, \lambda_2)$ we first change the integration variable to $x = \sqrt{a}\kappa$, so
that the new integration limits are $\pm \sqrt{a} \pi$. Since $\sqrt{a} \sim M$, in the thermodynamic limit the
integration interval becomes $(-\infty,\infty)$, leading to
\begin{equation}
\label{eq:strat_avg_m2} \overline{m}(\lambda_1, \lambda_2) = {N p_{0}^{2} \over \sqrt{8\pi a}} e^{-a\|\lambda_1 - \lambda_2 \|^{2}/2}.
\end{equation}
Inserting the expression for $\overline{m}(\lambda_1, \lambda_2)$ and $\overline{\ell}_{nn}(\kappa)$ into
Eqs.~(\ref{eq:cBi_final}) and (\ref{eq:avg_cBi_k}) yields the average bipartite clustering coefficient:
\begin{equation}
\label{eq:strat_cBi} \overline{c_{B}}(k) = \overline{c_{B}}(\kappa) = {1 \over 3}.
\end{equation}

To validate the obtained analytical expressions we perform numerical simulations, generating
networks with $N=10^{5}$ and $M=2\times10^{5}$.
To generate a network with a target value of $\overline{k}$ we set $\widetilde{a}$ according to
Eq.~(\ref{eq:strat_k2}). Figure~\ref{fig:strat}(a) shows the degree distributions for the top and bottom nodes in the
model. The degree distributions are well approximated by the Poisson distributions with the averages at $\overline{k} =
20$ and $\overline{\ell} = \overline{k} {N / M}$. Figure~\ref{fig:strat}(b) confirms that there are no correlations:
$\overline{\ell}_{nn}(k)$ and $\overline{k}_{nn}(\ell)$ do not depend on node degree, and match
Eq.~(\ref{eq:strat_lnn}).
Figure~\ref{fig:strat}(c) shows that clustering is strong, does not depend on either node degree or sizes $N,M$, and
matches the prediction in Eq.~(\ref{eq:strat_cBi}).
The appearance of high bipartite clustering in the stratified
model is due to preferential linking of nodes with similar hidden variables.

\section{Summary}

We have constructed and analyzed a general class of bipartite networks with hidden variables. In this class of bipartite
networks, nodes of both type reside in hidden variable spaces, and the connection
probability between a pair of nodes is a function of their hidden variables. The independent character of link appearance in
the model allows one to calculate analytical expressions for many important topological properties of modeled networks.

The formalism developed here builds up on the hidden variable formalism for unipartite
networks~\cite{boguna03}. Some basic structural properties of bipartite networks, such as the degree distributions and
correlations, are straightforward generalizations of those in unipartite networks. Some other
characteristics, such as unipartite projections and bipartite clustering, are unique to bipartite networks.

The hidden variable formalism has proven to be a powerful tool in studying the structure and function of complex
networks~\cite{garlaschelli07,garlaschelli08,fekete09,miller07}.
One particular application of interest for us are network geometry and
navigability~\cite{boguna09,boguna09b,krioukov10,boguna10}.
The formalism developed here can also be useful in inferring individual characteristics, attributes, and annotations of
nodes in real bipartite networks.

\begin{acknowledgments}
We thank F.~Papadopoulos, M.~{\'A}.~Serrano, M.~Bogu{\~n}{\'a} and kc claffy for many useful discussions and
suggestions. This work was supported by NSF Grants No.~CNS-0964236, CNS-1039646, CNS-0722070; DHS Grant
No.~N66001-08-C-2029; and by Cisco Systems.
\end{acknowledgments}

\begin{widetext}
\appendix
\section{Derivation of the bipartite clustering coefficient}
Here we provide the detailed derivation of the bipartite clustering coefficient defined in Eq.~(\ref{eq:cBi}).
We estimate the average bipartite clustering coefficient of a node with hidden variable $\kappa$ by calculating the ensemble
averages of the numerator and the denominator in Eq.~(\ref{eq:cBi}):
\begin{equation}
\label{eq:c42} \overline{c_{B}}(\kappa) =  { \langle \sum_{j > l}(m_{jl}-1) \rangle \over \langle \sum_{j > l}
(k_{j}+k_{l} - m_{jl} - 1) \rangle}.
\end{equation}
We first focus on the numerator in Eq.~(\ref{eq:c42}):
\begin{equation}
\label{eq:c4_numer} \langle \sum_{j > l}(m_{jl}-1) \rangle = {1 \over 2}\sum_{k} g(k|\kappa) k (k-1)
\sum_{\lambda_{1},\lambda_{2}}\rho(\lambda_{1}|\kappa)\rho(\lambda_{2}|\kappa)
\sum_{m}(m-1)P_{\lambda_{1},\lambda_{2}}(m-1),
\end{equation}
where $g(k|\kappa)$ is the $\kappa$-to-$k$ propagator, $\rho(\lambda_{1}|\kappa)$ is the conditional probability that a bottom node has hidden variable $\lambda_{1}$
provided it is connected to a top node with $\kappa$, and $P_{\lambda_{1},\lambda_{2}}(m-1)$ is the probability that
two bottom nodes with $\lambda_1$ and $\lambda_2$ have exactly $m-1$ common neighbors besides $i$.
Equation~(\ref{eq:c4_numer}) simplifies to
\begin{equation}
\label{eq:c4_numer2} \langle \sum_{j > l}(m_{jl}-1) \rangle = {1 \over 2} \langle k (k-1)\rangle_{\kappa}
\sum_{\lambda_{1},\lambda_{2}} P( \lambda_{1}|\kappa ) P(\lambda_{2}|\kappa) \overline{m}(\lambda_{1},\lambda_{2}).
\end{equation}
Next we compute the denominator of Eq.~(\ref{eq:c42}):
\begin{equation}
\label{eq:c4_denom} \langle \sum_{j > l} (k_{j}+k_{l} - m_{jl} - 1) \rangle = \langle \sum_{j > l} (k_{j}-1+k_{l}-1)
\rangle - \langle \sum_{j > l}(m_{jl}-1) \rangle.
\end{equation}
Sum $\langle \sum_{j > l}(m_{jl}-1) \rangle $ is the same as in the numerator, so that we only need to compute $\langle
\sum_{j > l} (k_{j}-1 + k_{l}-1) \rangle$:
\begin{equation}
\label{eq:c4_denom2} \sum_{j > l}(k_{j}-1 + k_{l}-1) = (k_{i}-1)\sum_{j=1}^{k_{i}}(k_{j}-1) = (k_{i}-1)k_{i}(k_{j}-1),
\end{equation}
where $k_{i}$ is degree of node $i$. Therefore,
\begin{equation}
\label{eq:c4_denom4} \langle \sum_{j > l} (k_{j}-1 + k_{l}-1) \rangle = \sum_{k} g(k|\kappa) (k-1)k
\sum_{\lambda}\rho(\lambda|\kappa)\sum_{\ell}(\ell-1)f(\ell-1|\lambda) = \langle k(k-1) \rangle_{\kappa}
\overline{\ell}_{nn}(\kappa).
\end{equation}
Using Eqs.~(\ref{eq:c4_numer2}) and (\ref{eq:c4_denom4}) we finally obtain
\begin{equation}
\label{eq:c4_fina} \overline{c_{B}}(\kappa) = {
\sum_{\lambda_1,\lambda_2}\rho(\lambda_{1}|\kappa)\rho(\lambda_{2}|\kappa) \overline{m}(\lambda_1,\lambda_2) \over 2
\overline{\ell}_{nn}(\kappa) - \sum_{\lambda_1,\lambda_2}\rho(\lambda_{1}|\kappa)\rho(\lambda_{2}|\kappa)
\overline{m}(\lambda_1,\lambda_2)}.
\end{equation}
\end{widetext}
\end{document}